\documentstyle[amssymb,12pt,psfig]{article}

\date{}
\begin{document}

\title{\textbf{BCS and BEC Finally Unified: \\ A Brief Review}}
\author{J. Batle,$^{a}$ M. Casas,$^{a}$ M. Fortes,$^{b}$ M. de Llano,$^{c}$\\
O. Rojo,$^{d}$ F.J. Sevilla,$^{b}$ M.A. Sol\'{\i}s$^{b}$ and 
V.V.Tolmachev$^{e}$ \\
$^{a}$Departament de F\'{\i}sica, Universitat de les Illes Balears,\\
07071 Palma de Mallorca, Spain\\
$^{b}$Instituto de F\'{\i}sica, UNAM, 01000 M\'{e}xico, DF, Mexico\\
$^{c}$Instituto de Investigaciones en Materiales, UNAM,\\
04510 M\'{e}xico, DF, Mexico\\
$^{d}$PESTIC, Secretar\'{\i}a Acad\'{e}mica, IPN, M\'{e}xico DF, Mexico\\
$^{e}$Baumann State Technical University, Moscow, Russia.}
\maketitle

\section{Introduction}

We review efforts to unify both the Bardeen, Cooper \& Schrieffer (BCS) and
Bose-Einstein condensation (BEC) pictures of superconductivity. We have
finally achieved this in terms of a ``\textit{complete} boson-fermion (BF)
model'' (CBFM) that reduces in special cases to all the main continuum (as
opposed to ``spin'') statistical theories of superconductivity. Our BF model
is ``complete'' in the sense that not only{\large \ }two-electron (2e) but
also two-hole (2h) Cooper pairs (CPs) are allowed in arbitrary proportions.
In contrast, BCS-Bogoliubov theory---which can also be considered as the
theory of a mixture of kinematically independent electrons, 2e- and
2h-CPs---allows only equal, 50\%-50\%, mixtures of the two kinds of CPs.
This is obvious from the perfect symmetry\ about $\mu $, the electron
chemical potential, of the well-known Bogoliubov \cite{Bog} $v^{2}(\epsilon )
$\ and $u^{2}(\epsilon )$ coefficients, where $\epsilon $ is the electron
energy. The CBFM is then applied to see: a) whether\ the BCS model
interaction for the electron-phonon dynamical mechanism is sufficient to
predict the unusually high values \cite{Uemura} of $T_{c}$ (in units of the
Fermi temperature) of $\simeq 0.01-0.1$\ exhibited by the so-called
``exotic'' superconductors \cite{Brandow}\ in both 2D and 3D---relative to
the low values of $\lesssim 10^{-3}$ more or less correctly predicted by BCS
theory for conventional, elemental superconductors; and b) whether it can at
least suggest, if not explain, why ``hole superconductors'' have higher $%
T_{c}$'s.

Boson-fermion (BF) models of superconductivity as a BEC \cite{Ogg,Ginz} go
back to the mid-1950's \cite{BF}-\cite{Blatt}, pre-dating even the
BCS-Bogoliubov theory \cite{bcs,bts}. Although BCS theory only contemplates
the presence of ``Cooper correlations'' of single-particle states, BF models 
%\cite{BF,bcs,bts,BF1,BF2,BF4,BF5,BF5a,BF5b,BF6,BF6a,BF7,BF7a,PLA2,BF8,FW,BF9}
\cite{BF}-\cite{Blatt}, \cite{BF3}-\cite{BF10} posit the existence of
actual bosonic CPs. Such pair charge carriers have been observed in magnetic
flux quantization experiments with elemental \cite{classical,classical2} as
well as with cuprate \cite{cuprates}\ superconductors. But apparently no
experiment has yet been done that{\large \ }distinguishes between electron
and hole CPs. The fundamental drawback of early \cite{BF}-\cite{Blatt} BF
models, which considered 2e bosons in analogy with diatomic molecules in a
classical gas mixture, is the notorious absence of an electron energy gap $%
\Delta $. The gap first began to appear in later BF models \cite{BF3}-\cite%
{PLA2}. With two \cite{BF7a,PLA2}\ exceptions, however, all BF models
neglect the effect of \textit{hole }CPs formulated on an equal footing with
electron CPs to give\ us a complete BF model (CBFM) consisting of \textit{%
both} bosonic CP species coexisting with \textit{unpaired} electrons.

Without going into a detailed justification we merely list several common
``myths'' in the theory of superconductivity that we tacitly \textit{%
disbelieve}: 

\begin{enumerate}
\item With the electron-phonon dynamical mechanism transition
temperatures $T_{c}\lesssim 45$ K at most. For higher $T_{c}$'s\ one needs
magnons or excitons or\ plasmons or other electronic mechanisms.

\item Cooper pairs (CPs):

\begin{description}
\item[a)] consist of negative-energy stable (i.e., stationary) bound
states \cite{Coo};

\item[b)] propagate in the Fermi sea with energy $\hbar ^{2}K^{2}/2(2m)$
(Ref. \cite{Blatt},\ p. 94) where $\hbar K$ is the total or center-of-mass
momentum (CMM) of the composite pair;

\item[c)] have a linear dispersion $E \varpropto v_{F}\hbar K$, with 
$v_{F}$ the Fermi velocity, which is merely the acoustic mode in the ideal
Fermi gas (+ interactions) with sound speed $v_{F}/\sqrt{d}$ in any
dimensionality $d$;

\item[d)] ``...with $K\neq 0$ represent states with net current flow \cite%
{NobelLecture};''

\item[e)] are \textit{not} bosons (Ref. \cite{Schrieffer}, p. 38). And
most notoriously, that: \ 
\end{description}
\item Superconductivity is unrelated to BEC \cite{Bardeen}.
\end{enumerate}
We question all of these assertions which will be discussed in greater
detail elsewhere.

\section{The CBFM Hamiltonian}

The CBFM \cite{BF7a,PLA2} is described by $H=H_{0}+H_{int}$ where the
unperturbed Hamiltonian $H_{0}$\ corresponds to an \textit{ideal }(i.e.,
noninteracting) gas mixture of fermions and both types of CPs, two-electron
(2e) and two-hole (2h), namely 
\begin{equation}
H_{0}=\sum\limits_{\mathbf{k}_{1},s_{1}}\varepsilon _{\mathbf{k}_{\mathbf{1}%
}}a_{\mathbf{k}_{1},s_{_{1}}}^{+}a_{\mathbf{k}_{1},s_{_{1}}}+\sum\limits_{%
\mathbf{K}}E_{+}(K)b_{\mathbf{K}}^{+}b_{\mathbf{K}}-\sum\limits_{\mathbf{K}%
}E_{-}(K)c_{\mathbf{K}}^{+}c_{\mathbf{K}},  \label{H0}
\end{equation}
where $\mathbf{K\equiv k}_{\mathbf{1}}+\mathbf{k}_{\mathbf{2}}$ is the CMM
wavevector, $\mathbf{k}\equiv \frac{1}{2}\mathbf{(k}_{\mathbf{1}}-\mathbf{k}%
_{\mathbf{2}})$ being the relative one, while $\varepsilon _{\mathbf{k}%
}\equiv \hbar ^{2}k^{2}/2m$ are the electron and $E_{\pm }(K)$\ the
2e-/2h-CP energies.\ Here $a_{\mathbf{k}_{1},s_{1}}^{+}$ ($a_{\mathbf{k}%
_{1},s_{1}}$) are creation (annihilation) operators for fermions and
similarly $b_{\mathbf{K}}^{+}$ ($b_{\mathbf{K}}$) and $c_{\mathbf{K}}^{+}$ ($%
c_{\mathbf{K}}$) for 2e- and 2h-CP bosons, respectively.

Two-hole CPs are considered \textit{distinct }and\textit{\ kinematically
independent }from 2e-CPs as their Bose commutation relations involve a
relative sign change, in sharp contrast with electron or hole fermions whose
Fermi anticommutation relations do not.\ In fact, holes have a dramatic
effect even in the simple, elementary CP problem where they were originally
neglected thereby giving \cite{Coo} a \textit{negative-}real-energy,
stationary (i.e., infinite-lifetime) two-fermion bound-state. But if
electrons and holes are treated equally and simultaneously through a
Bethe-Salpeter (BS) equation (see, e.g., Ref. \cite{FW} p. 131) in the ideal
Fermi gas (IFG) ground-state about which the CPs are defined, the resulting
energy is pure \textit{imaginary} \cite{bts},\cite{AGD}---implying an
obvious instability. The IFG-based CP problem is thus meaningless if
particles are taken on an equal footing with holes, as consistency would
demand. However, a similar BS treatment not about the IFG but about the BCS
ground-state yields \cite{Honolulu} real (but \textit{positive}, as with a
``quasi-bound state in the continuum'') 2e- \textit{and} 2h-CP energies,
along with an imaginary part that is nonzero only for $K\neq 0$ signifying a
finite lifetime, but zero for $K=0$ implying permanent pairs. Thus, the CP
problem is vindicated in a very natural, physical way via the BS equation.

The interaction Hamiltonian $H_{int}$ consists of four distinct interaction
vertices, each with two-fermion/one-boson creation or annihilation
operators, depicting how unpaired electrons (subindex +) [or holes (subindex 
$-$)] combine to form the 2e- (and 2h-CPs) assumed in the system of size $L$%
, namely 
\[
H_{int}=L^{-3/2}\sum\limits_{\mathbf{k},\mathbf{K}}f_{+}(k)\{a_{\mathbf{k}+%
\frac{1}{2}\mathbf{K},\uparrow }^{+}a_{-\mathbf{k}+\frac{1}{2}\mathbf{K}%
,\downarrow }^{+}b_{\mathbf{K}} + a_{-\mathbf{k}+\frac{1}{2}%
\mathbf{K},\downarrow }a_{\mathbf{k}+\frac{1}{2}\mathbf{K},\uparrow }b_{%
\mathbf{K}}^{+}\}
\]
\begin{equation}
+L^{-3/2}\sum\limits_{\mathbf{k},\mathbf{K}}f_{-}(k)\{a_{\mathbf{k}+\frac{1}{%
2}\mathbf{K},\uparrow }^{+}a_{-\mathbf{k}+\frac{1}{2}\mathbf{K},\downarrow
}^{+}c_{\mathbf{K}}^{+} + a_{-\mathbf{k}+\frac{1}{2}\mathbf{K}%
,\downarrow }a_{\mathbf{k}+\frac{1}{2}\mathbf{K},\uparrow }c_{\mathbf{K}}\}.
\label{Hint}
\end{equation}%
Note that the \textit{fermion-pair interaction} $H_{int}$ is reminiscent%
\textit{\ }of\textit{\ }the Fr\"{o}hlich (or Dirac\ QED) interaction
Hamiltonian involving two fermion and one boson operators but with \textit{%
two}\textbf{\ }types of CPs instead of phonons (or photons). But in contrast
with Fr\"{o}hlich and Dirac there is no a conservation law for the number of
unpaired electrons, i.e.,$\ [H_{int},\sum\limits_{\mathbf{k}%
_{1},s_{1}}\varepsilon _{\mathbf{k}_{\mathbf{1}}}a_{\mathbf{k}%
_{1},s_{_{1}}}^{+}a_{\mathbf{k}_{1},s_{_{1}}}]\neq 0.$ (Note too that{\Huge 
{\large \ }}$[H_{int},\sum\limits_{\mathbf{k}_{1},s_{1}}\mathbf{k}_{1}a_{%
\mathbf{k}_{1},s_{_{1}}}^{+}a_{\mathbf{k}_{1},s_{_{1}}}]=0$ and $%
[H_{int},\sum\limits_{\mathbf{k}_{1},s_{1}}s_{1}a_{\mathbf{k}%
_{1},s_{_{1}}}^{+}a_{\mathbf{k}_{1},s_{_{1}}}]=0$.){\Huge \ }Just as the Fr%
\"{o}hlich (or Dirac) interaction Hamiltonians are the most natural ones to
use in a many-electron/phonon (or photon) system, one can conjecture the
same of (\ref{Hint}) for the BF system under study. Indeed, this{\large \ }$%
H_{int}$ has \textit{formally} already been employed under various guises by
several authors \cite{BF3},\cite{BF4}-\cite{BF8}. The energy form factors $%
f_{\pm }(k)$ are\ essentially the Fourier transforms of the 2e- and 2h-CP
intrinsic wavefunctions, respectively, in the relative coordinate between
the paired fermions of the CP. Here they are taken simply as
\begin{equation}
f_{+}(\varepsilon )=\left\{ 
\begin{array}{cc}
f & \quad {\hbox{for}} \,\,E_{f}<\varepsilon <E_{f}+\delta \varepsilon ,\quad
\\ 
0 & {\hbox{otherwise,}}%
\end{array}%
\right.  \label{f+}
\end{equation}%
\begin{equation}
f_{-}(\varepsilon )=\left\{ 
\begin{array}{cc}
f & \quad {\hbox{for}}\,\,E_{f}-\delta \varepsilon <\varepsilon <E_{f},\quad
\\ 
0 & {\hbox{otherwise,}}%
\end{array}%
\right.  \label{f-}
\end{equation}%
with $E_{f}$ and $\delta \varepsilon $ phenomenological dynamical energy
parameters (in addition to the positive coupling parameter $f$) related to
the bosonic CPs through $E_{f}\equiv \frac{1}{4}[E_{+}(0)+E_{-}(0)]$ and $%
\delta \varepsilon \,\equiv \,\frac{1}{2}[E_{+}(0)-E_{-}(0)]$, where $%
E_{\pm }(0)$ are the (empirically \textit{un}known) zero-CMM energies of the
2e- and 2h-CPs, respectively. Clearly $E_{\pm }(0)=2E_{f}\pm \delta
\varepsilon $. The quantity\ $E_{f}$ will serve as a convenient energy scale
and is not to be confused with the Fermi energy $E_{F}=\frac{1}{2}%
mv_{F}^{2}\equiv k_{B}T_{F}$ where $T_{F}$\ is the Fermi temperature. The
Fermi energy $E_{F}$ equals $\pi \hbar ^{2}n/m$ in 2D and $(\hbar
^{2}/2m)(3\pi ^{2}n)^{2/3}$ in 3D, with $n$ the total number-density of
charge-carrier electrons. The quantities$\ E_{f}$ and $E_{F}$ coincide 
\textit{only }when perfect 2e/2h-CP symmetry holds.

%FIGURE 1
\begin{figure}[tbh]
\centerline{\hspace{-2.0cm}\psfig{file=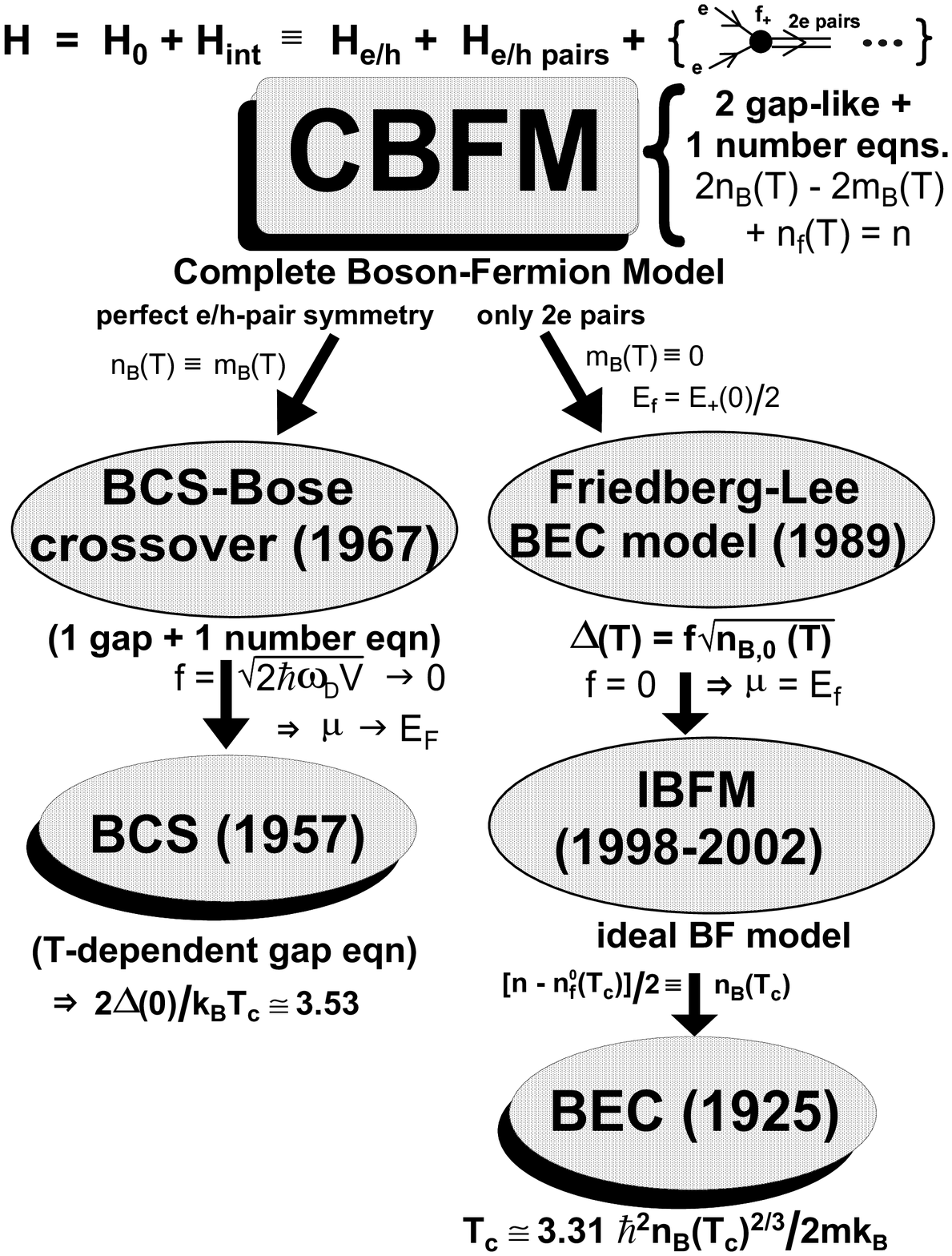,height=8.50in,width=7.0in}}
%\vspace{-2.0cm}
\caption{Flow chart of how the CBFM reduces in special cases to the statistical,
continuum models of superconductivity discussed in text, thereby displaying
how both BCS and BEC theories can be unified.}
\end{figure}

\section{Main statistical theories as special cases of CBFM}

The interaction Hamiltonian (\ref{Hint}) can be further simplified by
keeping only the $\mathbf{K}=0$ terms. One then applies the Bogoliubov
``recipe'' (see, e.g., \cite{FW} p. 199) of replacing in the full
hamiltonian $H=H_{0}+H_{int}$ all zero CMM creation and annihilation
operators by c-numbers: $\sqrt{N_{0}}$ and $\sqrt{M_{0}}$ for 2e- and 2h-CP
operators, where $N_{0}(T)$ and $M_{0}(T)$ are the number of zero-CMM 2e-
and 2h-CPs, respectively. Minimizing with respect to the independent
variables $N_{0}$\ and $M_{0}$ the so-called thermodynamic (or grand)
potential associated with the full Hamiltonian$\ H$ , as well as keeping the
total number of electrons fixed and thereby introducing the electron
chemical potential $\mu $, yields a set of three coupled, transcendental,
integral equations (Ref. \cite{BF7a}, Eqs. 7-8). These three equations
embody the CBFM. Two of these are coupled gap-like equations involving the
2e-CP and 2h-CP BE-condensed boson number densities $n_{0}(T)\equiv
N_{0}(T)/L^{d}$ and $m_{0}(T)\equiv M_{0}(T)/L^{d}$, linked together through
an electron energy gap $\Delta $. The third equation can be cast as a number
equation of the form $2n_{B}(T)-2m_{B}(T)+n_{f}(T)=n$\ involving both 2e and
2h boson\textit{\ }number-densities but now for all energy states, where $%
n_{f}(T)$ is the number-density of unpaired\textit{\ }electrons. Most
significantly, these three equations contain \textit{five }different
theories as special cases, see flow chart in Fig. 1. For perfect 2e/2h CP
symmetry $n_{B}(T)=m_{B}(T)$ implies \cite{BF7a}\ that $n_{0}(T)=m_{0}(T)$
and that $E_{f}$ coincides with $\mu $. The CBFM then reduces to: \textbf{i)}
the gap and number equations of the \textit{BCS-Bose crossover picture} \cite%
{BCS-Bose} for the BCS model interaction---if the BCS parameters $V$ and
Debye energy $\hbar \,\omega _{D}$ are properly identified with the CBFM
dynamical parameters through $f^{2}/2\delta \varepsilon $ and $\delta
\varepsilon $, respectively. The crossover picture for unknowns $\Delta (T)$
and $\mu (T)$ is now supplemented by the key relation $\Delta (T_c)=f\sqrt{%
n_{0}(T_c)}=f\sqrt{m_{0}(T_c)}$. If in addition one imposes that $\mu \simeq
E_{F}$, as occurs for weak coupling from the number equation,\ the crossover
picture is well-known to reduce to: \textbf{ii)} \textit{ordinary BCS theory}%
. Thus, the BCS condensate is \textit{precisely} a BE condensate whenever $%
n_{B}(T_c)=m_{B}(T_c)$ and the coupling is weak.

On the other hand, for no 2h-CPs present the CBFM reduces \cite{BF7a}\ also
to: \textbf{iii)} the \textit{BEC BF model} in 3D of Friedberg and Lee \cite%
{BF5,BF6} characterized by the relation $\Delta (T)=f\sqrt{n_{0}(T)}$. With
just \textit{one} adjustable parameter (the ratio of perpendicular to planar
boson masses) this theory fitted \cite{BF6}\ cuprate $T_{c}/T_{F}$ data
quite well. When $f=0$ the CBFM reduces\ to both: \textbf{iv)} the ideal BF
model of Ref. \cite{BF9,BF10} that predicts nonzero BEC $T_{c}$'s even in
2D,\ as well as to: \textbf{v)} the familiar $T_{c}$-formula of ordinary BEC
in 3D, albeit as an \textit{implicit }equation with the boson number-density
being $T$-dependent. Figure 2 displays the $T_{c}$ prediction \cite{BF10} in
2D for cuprate superconductors of special case (iv), with \textit{no}
adjustable parameters.

\section{BEC limit of all electrons paired}

The general BEC $T_{c}$-formula for noninteracting bosons in $d$-dimensions
of energy $\varepsilon _{K}=C_{s}\,K^{s},$ $s>0,$ is \cite{B8a}
\begin{equation}
T_{c}=\frac{C_{s}}{k_{B}}\left[ \frac{s\Gamma (d/2)(2\pi )^{d}}{2\pi
^{d/2}\Gamma (d/s)g_{d/s}(1)}n_{B}\right] ^{s/d}  \label{Tc}
\end{equation}%
where $n_{B}$ is the boson number-density and the Bose
integral \cite{Path} $g_{\sigma}(z)\equiv \sum\limits_{l{\bf =1}}^{\infty
}z^{l}/l^{\sigma }$, with $z\equiv e^{\mu _{B}/k_{B}T}$ is the ``%
\textit{fugacity}''\ and $\mu _{B}$ the boson chemical potential. For $z=1$, 
$g_{\sigma }(1)$ $\equiv $ $\zeta (\sigma )$, the Riemann Zeta-function, if $%
\sigma >1$, while for $0<\sigma \leq 1$ the infinite series $g_{\sigma }(1)$
diverges.\ Eq. (\ref{Tc}) is formally valid for all $d>0$ and $s>0$. Hence,
for $0<d\leq s$, $T_{c}=0$ since $g_{d/s}(1)=\infty $ for $d/s\leq 1$ but $%
T_{c}$ is otherwise finite. We stress that as a consequence of the former 
\textit{all }2D\textit{\ }$T_{c}$ predictions in Fig. 2 (except the BCS one
that survives for all $d>0$) would collapse to zero had $s=2$ been used in
2D instead of the correct $s=1$. For $s=2$, $C_{2}=\hbar ^{2}/2m_{B}$ (\ref%
{Tc})\ leads to the familiar 3D result $T_{c}\simeq {3.31\hbar
^{2}n_{B}^{2/3}/}m_{B}k_{B}$ since $\zeta (3/2)\simeq 2.612$. Recalling that 
$k_{B}T_{F}=$ $\hbar ^{2}k_{F}^{2}/2m$ with $k_{F}=[2^{d-2}\pi ^{d/2}d\Gamma
(d/2)n]^{1/d}$, then if $m_{B}=2m$ and $n_{B}=n/2$ (all electrons paired)
for $s=2$ (\ref{Tc}) gives $T_{c}/T_{F}=\frac{1}{2}[2/d\Gamma
(d)g_{d/2}(1)]^{2/d}$ $=0$ for $d\leq 2$ since $g_{d/2}(1)=\infty $ for $%
d/2\leq 1$. For $d=3$ we arrive at another familiar result $T_{c}/T_{F}=%
\frac{1}{2}\left[ 2/3\Gamma (3/2)\zeta (3/2)\right] \simeq 0.218$ (see
dashed line in ``Uemura plot'' of Ref. \cite{Uemura}, Fig. 2). This value
appears marked in Fig. 4 as a black triangle.

%FIGURE 2
\begin{figure}[tbh]
\centerline{\psfig{file=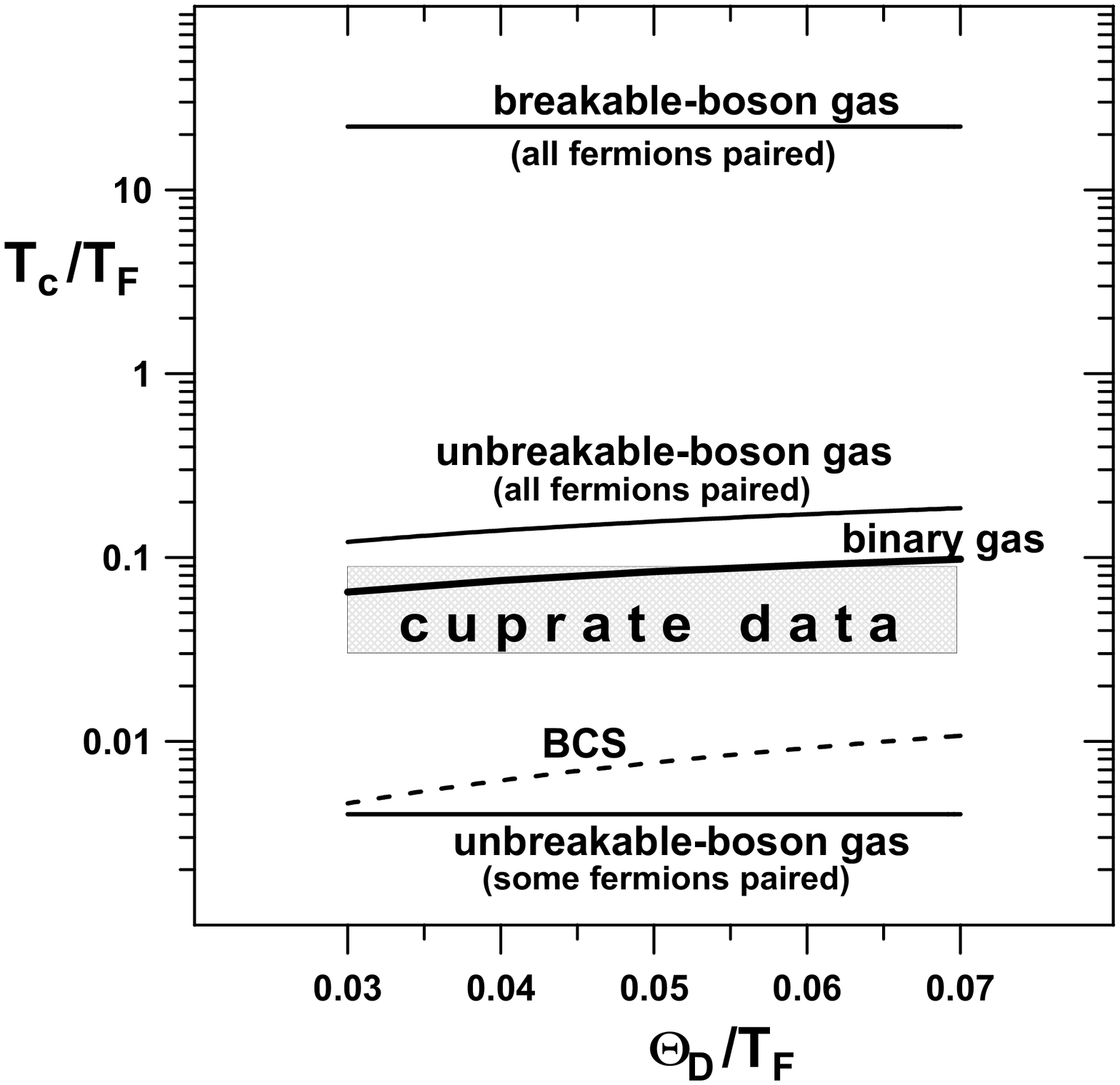,height=4.0in,width=4.0in}}
%\vspace{-2.0cm}
\caption{Critical 2D BEC-like temperature $T_{c}$ in units of $T_{F}$ for the BCS
model interaction with $\lambda =1/2$ for varying $\hbar \omega
_{D}/E_{F}\equiv \Theta _{D}/T_{F}$ for: the pure unbreakable-boson gas with 
\textit{some} and with \textit{all} fermions paired; for the breakable-boson
gas; and for the boson-fermion mixture (thick full curve labeled ``binary
gas'') in thermal/chemical equilibrium, all as described in Ref. \cite{BF10}
for the original (simple) CPs where $C_{1}=(2/\pi )\hbar v_{F}$. Dashed
curve is the BCS theory $T_{c}$, and cuprate experimental data are taken
from Ref. \cite{Poole}.}
\end{figure}

We now focus on $s=1$.\ For the boson energy $\eta $ to be used below\ the
leading term in the many-body Bethe-Salpeter (BS) CP dispersion relation is 
\textit{linear}, i.e., $\eta \simeq (\lambda /2\pi )\hbar v_{F}K$ [see Ref. %
\cite{Honolulu} for the derivation in 3D which gives $\eta \simeq (\lambda
/4)\hbar v_{F}K$]. Here $\lambda \equiv VN(E_{F})$ where $N(E_{F})$ is the
electron density of states (DOS) (for one spin) at the Fermi surface. Note
that the boson energy $\eta $ is \textit{linear} in CMM $K$---and \textit{%
not }the quadratic $\hbar ^{2}K^{2}/4m$ appropriate for a composite boson of
mass $2m$ moving not in the Fermi sea but in vacuum \cite{BF}-\cite{Blatt}, %
\cite{BF5}-\cite{PLA2}. The quadratic holds only when $E_{F}$ is \textit{%
strictly} zero \cite{PRB2000}, i.e., when no Fermi sea is present. These
linearly-dispersive\ CPs are commonly confused with the also
linearly-dispersive sound phonons of the collective excitation sometimes
referred to as the Anderson-Bogoliubov-Higgs (ABH) (Ref. \cite{bts} Sec. 3; %
\cite{ABH,Higgs})\ mode, which for zero coupling reduces \cite{Ran} to the
IFG result $\hbar v_{F}K/\sqrt{d}$. The IFG sound speed $c=v_{F}/\sqrt{d}$
also follows directly from the zero-temperature IFG pressure $%
P=n^{2}[d(E/N)/dn]=2nE_{F}/(d+2)$ via the familiar thermodynamic relation $%
dP/dn=mc^{2}$, where $E$ is the ground-state energy and as before $n\equiv
N/L^{d}=k_{F}^{d}/2^{d-2}\pi ^{d/2}d\;\Gamma (d/2)$ is the fermion-number
density. But the above results with $\eta \propto \lambda \hbar v_{F}K$ in
fact refer to actual ``moving'' (or ``excited'') CPs \textit{in the Fermi
sea.} \textit{Both} kinds of \textit{distinct} soundwave-like
solutions---moving CPs and ABH phonons---appear in the many-body BS
ladder-summation scheme of Ref. \cite{Honolulu}. Note also that the BS CP
linear dispersion coefficient $\lambda /2\pi $ in 2D (or $\lambda /4$\ in
3D) contrasts markedly with the coupling-independent $2/\pi $ coefficient in
2D (or $1/2$ in 3D, as first quoted in Ref. \cite{Schrieffer}, p. 33)
obtained \cite{PC98}\ in the \textit{simple} CP problem \cite{Coo} which
ignores holes. Thus (\ref{Tc}) again with $n_{B}=n/2$, for $s=1$ and $%
C_{1}=\lambda b(d)\hbar v_{F}$ with $b(2)=1/2\pi $ and $b(3)=1/4$, yields $%
T_{c}/T_{F}=2\lambda b(d)/\left[ d\Gamma (d)\zeta (d)\right] ^{1/d}$ which
(for $\lambda =1/2$) is $\simeq 0.088$ if $d=2$ since $\zeta (2)=\pi ^{2}/6$%
, and $\simeq 0.129$ if $d=3$ since $\zeta (3)\simeq 1.202$.\ These two
values for $T_{c}/T_{F}$\ will appear as the uppermost black squares in Fig.
4 marking the BEC limiting values if \textit{all }electrons\ in our 2D or 3D
many-electron system were imagined paired into noninteracting bosons formed
with the BCS model interelectron interaction. For $\lambda =1/4$ the
lowermost black squares in Fig. 4 apply.

\section{Enhanced $T_{c}$'s from the CBFM}

We now apply this very general CBFM to exhibit the sizeable enhancements in $%
T_{c}$s over BCS theory that emerge for moderate departures from perfect
2e/2h-pair symmetry for the \textit{same }interaction model. The
pair-fermion interaction (\ref{Hint}) with (\ref{f+}) and (\ref{f-}) bears a
one-to-one correspondence with the more familiar ``direct'' interfermion
electron-phonon interaction, mimicked, e.g., in the BCS model interaction
(whose double Fourier transform is a negative constant $-V$ nonzero only
within an energy shell $2\hbar \omega _{D}$ about the Fermi surface, with $%
\omega _{D}$ the Debye frequency) if \cite{BF7a,PLA2} we set $%
f^{\,2}/2\delta \varepsilon \equiv V$ and $\delta \varepsilon \equiv \hbar
\omega _{D}$. The familiar dimensionless BCS model interaction parameters $%
\lambda \equiv N(E_{F})V$ and $\hbar \omega _{D}/E_{F}$ are then recovered.

The three coupled equations of the CBFM\ determining the $d$-dimensional
BE-condensate number-densities $n_{0}(T)$ and $m_{0}(T)$ of 2e- and 2h-CPs,
respectively, as well as the fermion chemical potential $\mu (T)$, were
solved numerically in 3D for $\lambda =1/5$ and $\hbar \,\omega _{D}/E_{F}=$ 
$0.001$ in Ref. \cite{PLA2} assuming a quadratic boson dispersion relation $%
\eta =\hbar ^{2}K^{2}/4m$.\ For this case Figure 3 maps the phase diagram in
the vicinity of the BCS $T_{c}$ value (marked BCS-B in the figure) at $%
\Delta n\equiv n/n_{f}-1=0$ (corresponding to perfect 2e/2h-CP symmetry)
where $n_{f}$ is the number of unpaired electrons for zero gap and zero
temperature, provided that $n_{f}\leq n$ \cite{BF7a,PLA2}. Besides the 
\textit{normal }phase ($n$) consisting of the ideal BF gas described by $%
H_{0}$, three different types of stable (plus several metastable, i.e., of
higher Helmholtz free energy) BEC-like phases emerged. These are two \textit{%
pure} phases of \textit{either} 2e- ($s+$)\ \textit{or} 2h-CP ($s-$)\
BE-condensates, and a lower temperature \textit{mixed }phase ($ss$) with
arbitrary proportions of 2e- \textit{and} 2h-CPs. Of greater physical
interest are the two higher-$T_{c}$ \textit{pure }phases so that we focus
below only on them. For each pure phase at a critical temperature we have 
\textit{either} $\Delta (T_{cs+})=f\sqrt{n_{0}(T_{cs+})}$ $\equiv 0$ or $%
\Delta (T_{cs-})=f\sqrt{m_{0}(T_{cs-})}\equiv 0$, where $\Delta (T)$ is the
electronic (BCS-like) energy gap. Their intersection gives the BCS $T_{c}$
value of $7.64\times 10^{-6}T_{f}$ where $k_{B}T_{f}\equiv E_{f}=(\hbar
^{2}/2m)(3\pi ^{2}n_{f})^{2/3}$.

\subsection{Two dimensions (2D)}

In 2D the one-spin electronic density of states (DOS) is constant, namely $%
N(\varepsilon )=m/2\pi \hbar ^{2}$. Using the aforementioned BS CP \textit{%
linear }dispersion relation $\eta \simeq (\lambda /2\pi )\hbar v_{F}K$ we
get for the bosonic DOS $M(\eta )\equiv (1/2\pi )K(dK/d\eta )\simeq (2\pi
/\lambda ^{2}\hbar ^{2}v_{F}^{2})\eta $ instead of the constant that follows
in 2D from quadratic dispersion. Employing $E_{f}\equiv \pi \hbar
^{2}n_{f}/m=k_{B}T_{f}$ as energy/density/temperature scaling factors, and
the relation $n/n_{f}=(E_{F}/E_{f})^{d/2}$ to convert quantities such as $%
T_{c}/T_{f}$ to $T_{c}/T_{F}$, where $E_{F}\equiv $ $k_{B}T_{F}$, the two
working equations for the \textit{pure 2e-CP phase }[i.e., $%
m_{B}(T_{c})\equiv 0$] become (with all quantities dimensionless, energies
in units of $E_{f}$ and electron particle-densities in units of $n_{f}$)
\begin{equation}
1+\hbar \omega_{D}/2-\mu =\lambda (\hbar \omega
_{D}/2)\int\limits_{1}^{1+\hbar \omega_{D}} dx \frac{1}{|x-\mu |} {%
\tanh }\frac{|x-\mu |}{2T_{c}},  \label{mu1}
\end{equation}%
\begin{equation}
\frac{1}{2}\int\limits_{0}^{\infty }dx[1- {\tanh }\frac{x-\mu }{2T_{c}}%
]+\frac{\pi ^{2}}{\lambda ^{2}}\frac{1}{n}\int\limits_{0}^{\infty }dxx[%
 {\coth }\,\frac{x+2(1+\hbar \omega _{D}/2-\mu )}{2T_{c}}-1]=n.
\label{n1}
\end{equation}%
These are just the gap-like equation associated with 2e-CPs and its
corresponding number equation. 

%FIGURE 3
\begin{figure}[tbh]
\centerline{\psfig{file=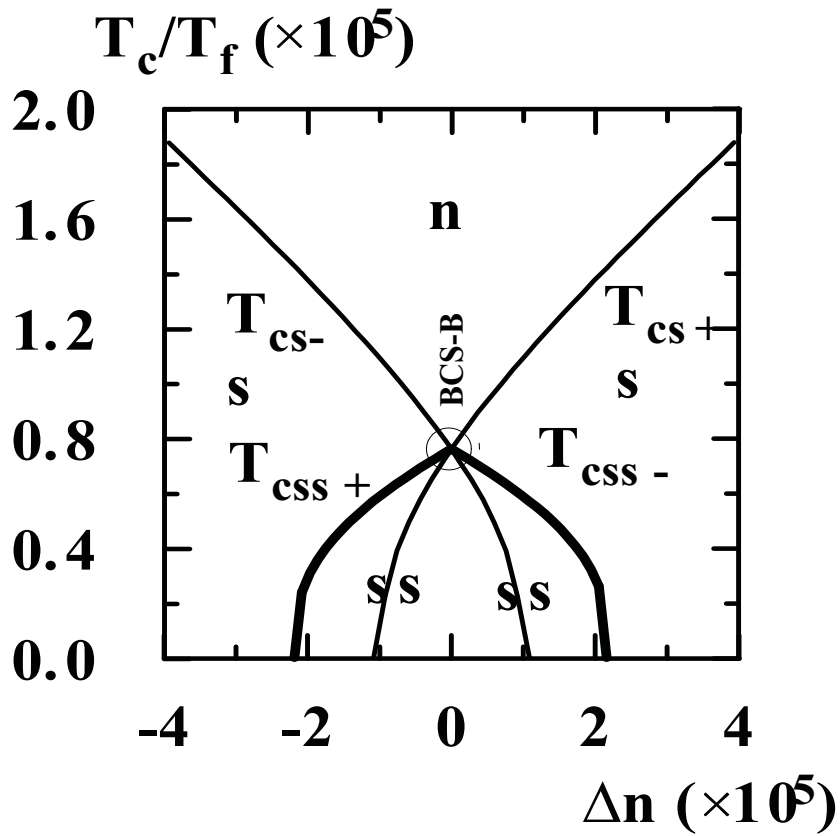,height=4.50in,width=4.50in}}
%\vspace{-2.0cm}
\caption{Phase diagram \cite{PLA2} with 3D superconducting critical temperatures $%
T_{cs+}$, $T_{cs-}$, $T_{css+}$, and $T_{css-}$ as functions of $\Delta
n\equiv n/n_{f}-1$ as defined in text,\ in the vicinity of the BCS $T_{c}$
value, assuming a quadratic boson dispersion, for $\lambda =1/5$ and $\hbar
\,\omega _{D}/E_{F}=$ $0.001$.}
\end{figure}

For the \textit{pure 2h-CP phase\ }(i.e., $n_{B}\equiv 0$) the two working
equations are
\begin{equation}
\mu -1+\hbar \omega _{D}/2=\lambda (\hbar \omega _{D}/2)\int\limits_{1-\hbar
\omega _{D}}^{1}dx\frac{1}{|x-\mu |} {\tanh }\frac{|x-\mu |}{2T_{c}},
\label{mu2}
\end{equation}
\begin{equation}
\frac{1}{2}\int\limits_{0}^{\infty }dx[1- {\tanh }\frac{x-\mu }{2T_{c}}%
]-\frac{\pi ^{2}}{\lambda ^{2}}\frac{1}{n}\int\limits_{0}^{\infty }dxx[%
 {\coth }\frac{x-2(1-\hbar \omega _{D}/2-\mu )}{2T_{c}}-1]=n.
\label{n2}
\end{equation}%
which are the gap-like equation associated with 2h-CPs and its corresponding
number equation. Note that (\ref{n1}) and (\ref{n2}) are quadratic in $n$,
and that all integrals there are exact, namely
\begin{equation}
\int\limits_{0}^{\infty }dx[1- {\tanh }\frac{x-\mu }{2T_{c}}]=\mu
+2T_{c} {\ln}[2 {\cosh}({\mu /2T}_{c})],  \label{mu3}
\end{equation}
and
\begin{equation}
\int\limits_{0}^{\infty }dxx[ {\coth }\frac{x+\delta }{2T_{c}}%
-1]=2T_{c}^{2} g_{2}(e^{-\delta /T_{c}})  \label{n3}
\end{equation}%
where the Bose function $g_{\sigma}(z)$ was defined after (5); it 
is designated in Ref. \cite{Wolfram} as PolyLog$[\sigma ,z]$. 
The integrals in (\ref{mu1}) and (\ref{mu2}) were
performed numerically. In 2D we use the two extreme values of $\lambda =1/4$
(lower set of curves in Fig. 4) and $=1/2$ (upper set of curves), and $\hbar
\omega _{D}/E_{F}=0.05$ (a typical value for cuprates), to compute from
equations (\ref{mu1}) to (\ref{n2}) the $T_{c}/T_{F}$ vs. $n/n_{f}$ phase
diagram which is graphed in the figure (left panel for 2D) for both 2e-CP
(dashed curve) and 2h-CP (full curve) pure, stable BEC-like phases. The
value $n/n_{f}=1$ corresponds to perfect 2e/2h-CP symmetry. The $T_{c}$
value where both curves $n_{0}(T_{c})=m_{0}(T_{c})$ $=0$ intersect\ is
marked by the large dots in the figure; these values are consistent with
those gotten from the familiar BCS expression $T_{c}/T_{F}\simeq 1.134(\hbar
\omega _{D}/E_{F})\exp (-1/\lambda )\simeq 0.001$ for $\lambda =1/4$, and $%
0.008$ for $\lambda =1/2$, for $\hbar \omega _{D}/E_{F}=0.05$. [Thes values
differ little from those from the exact BCS (implicit) $T_{c}$-formula (Ref. %
\cite{FW}, p. 447) $1$ $=\lambda \int_{0}^{\hbar \omega
_{D}/2k_{B}T_{c}}dxx^{-1}\tanh x$.] Cuprate data empirically \cite{Poole}
fall within the range $T_{c}/T_{F}\simeq 0.03-0.09$. Thus, moderate
departures from perfect 2e/2h-CP symmetry enable the CBFM to reach quasi-2D
cuprate empirical $T_{c}$ values, and quite likely also room temperature
superconductivity, \textit{without abandoning electron-phonon dynamics}%
---contrary to popular belief. Compelling evidence for a strong, if not
sole, phonon dynamical component in cuprates has recently been reported \cite%
{Shen} from angle-resolved-photoemission data.

\subsection{Three dimensions (3D)}

In 3D $N(\varepsilon )\equiv (1/2\pi ^{2})k^{2}(dk/d\varepsilon )=(m^{\frac{3%
}{2}}/2^{\frac{1}{2}}\pi ^{2}\hbar ^{3})\,\sqrt{\varepsilon }$ and,
analogously as before, $E_{f}=(\hbar^{2}/2m)\times$ $(3\pi ^{2}n_{f})^{2/3}\equiv
k_{B}T_{f}$ which again differs from $E_{F}=(\hbar ^{2}/2m)(3\pi
^{2}n)^{2/3}\equiv k_{B}T_{F}$ except when perfect 2e/2h-CP symmetry holds
when they coincide, whereas the leading term in the BS CP boson dispersion
energy is now the linear expression $\eta \simeq (\lambda /4)\hbar v_{F}K$ %
\cite{Honolulu} so that $M(\eta )\equiv (1/2\pi ^{2})K^{2}(dK/d\eta )\simeq
(32/\pi ^{2}\lambda ^{3}\hbar ^{3}v_{F}^{3})\eta ^{2}$. The above working
equations for the \textit{pure 2e-CP phase} now in 3D (all quantities again
dimensionless) are
%\begin{center}
\begin{equation}
1+\hbar \omega _{D}/2-\mu =\lambda (\hbar \omega _{D}/2)\frac{1}{n^{1/3}}%
\int\limits_{1}^{1+\hbar \omega _{D}}dx\sqrt{x}\frac{1}{|x-\mu |} {%
\tanh }\frac{|x-\mu |}{2T_{c}},  \label{mu3De}
\end{equation}%
%\end{center}
\begin{equation}
\frac{3}{4}\int\limits_{0}^{\infty }dx\sqrt{x}[1- {\tanh }\frac{x-\mu 
}{2T_{c}}]+\frac{12}{\lambda ^{3}}\frac{1}{n}\int\limits_{0}^{\infty
}dxx^{2}[ {\coth }\frac{x+2(1+\hbar \omega _{D}/2-\mu )}{2T_{c}}-1]=n,
\label{n3De}
\end{equation}%
while for the\textit{\ pure 2h-CP phase} they are
\begin{equation}
\mu -1+\hbar \omega _{D}/2=\lambda (\hbar \omega _{D}/2)\frac{1}{n^{1/3}}%
\int\limits_{1-\hbar \omega _{D}}^{1}dx\sqrt{x}\frac{1}{|x-\mu |} {%
\tanh }\frac{|x-\mu |}{2T_{c}},  \label{mu3Dh}
\end{equation}
\begin{equation}
\frac{3}{4}\int\limits_{0}^{\infty }dx\sqrt{x}[1- {\tanh }\frac{x-\mu 
}{2T_{c}}]-\frac{12}{\lambda ^{3}}\frac{1}{n}\int\limits_{0}^{\infty
}dxx^{2}[ {\coth }\frac{x-2(1-\hbar \omega _{D}/2-\mu )}{2T_{c}}-1]=n.
\label{n3Dh}
\end{equation}%
Results in 3D are reported only for the 2e-CP BEC case for the same extreme
values of $\lambda =1/4$ and $1/2$ as in 2D but now for $\hbar \omega
_{D}/E_{F}=0.005$. In Fig. 4 for 3D (right panel) the dashed curves are the
2e-CP BEC phase boundaries. The large dot again marks the BCS $T_{c}/T_{F}$
values of $0.0001$ for $\lambda =1/4$ and $0.0008$ for $\lambda =1/2$. 
%while the open circles again represent the $%
%T_{c}/T_{F}$ values above which the quadratic-in-$n/n_{f}$ expression\ (\ref%
%{n3Dh}) yields a complex solution. 
Empirical data for both exotic and conventional, elemental superconductors
in 3D are taken from Ref. \cite{Uemura}. We see that whereas BCS theory can
roughly reproduce $T_{c}/T_{F}$ values\ well for the latter, it takes
moderate departures from perfect 2e/2h-CP symmetry to access 3D exotic
superconductor $T_{c}/T_{F}$ values, which empirically \cite{Uemura} fall
within the range $\simeq 0.01-0.1$. This is much larger than the range $%
\lesssim 0.001$ for conventional (elemental) superconductors, also shaded in
the right panel of the figure. 

%FIGURE 4
\begin{figure}[tbh]
\centerline{\psfig{file=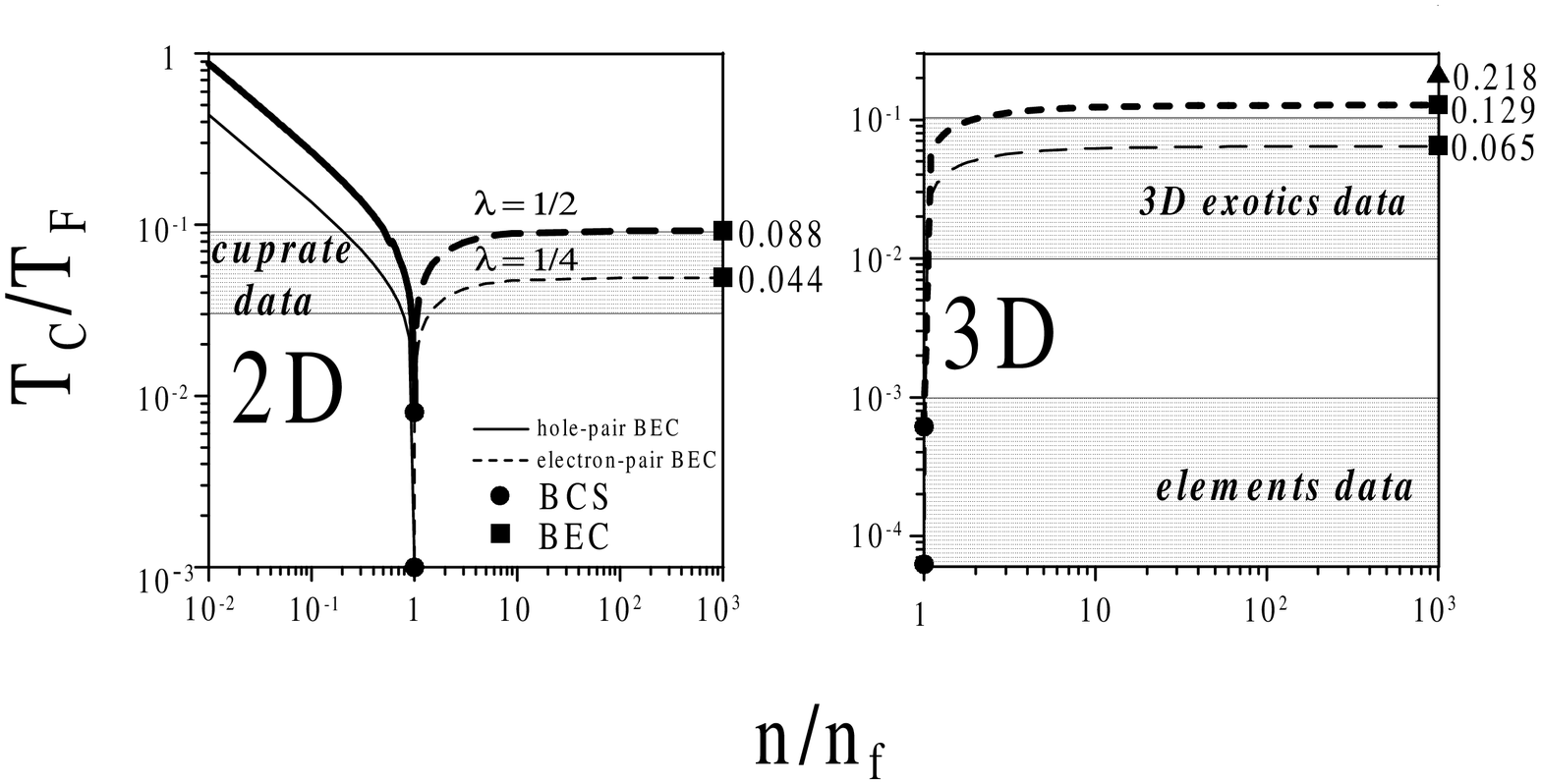,height=3.5in,width=7.0in}}
%\vspace{-2.0cm}
\caption{Phase diagrams in 2D and 3D for temperature (in units of $T_{F}$) and
electron density (in units of $n_{f}$ as defined in text) showing the phase
boundaries of $T_{c}$'s for the pure 2e-CP BEC phases (dashed curves)
determined by $\Delta (T_{c})=f\sqrt{n_{0}(T_{c})}$ $\equiv 0$, and the pure
2h-CP BEC phase (in 2D only) given by $\Delta (T_{c})=f\sqrt{m_{0}(T_{c})}%
\equiv 0$ for $\lambda =1/4$ and $1/2$ with $\hbar \omega _{D}/E_{F}=0.05$
in 2D and $0.005$ in 3D. Intersections corresponding to $%
n_{0}(T_{c})=m_{0}(T_{c})$ giving the BCS $T_{c}$ approximately are marked by black dots,
while black squares mark the BEC limit where all electrons are imagined
paired into 2e-CP bosons, and the black triangle marks
the familiar 3D limit as determined in Sec. 4.}
\end{figure}

\section{Hole superconductivity}

Finally, we address the unique but mysterious role played by \textit{holes}
in superconductors in general. For example: a) of the cuprates those that
are hole-doped have transition temperatures $T_{c}$ about \textit{six }times
higher than electron-doped ones; and b) in fullerite (an fcc crystal of $%
C_{60}$ fullerenes) $T_{c}$ is now claimed to be more than \textit{three}
times higher with hole rather than electron doping, as recently observed %
\cite{Batlogg}\ with the so-called ``field-effect transistor'' technique of
injecting holes. And even in conventional superconductors \cite{Hirsch} c)
over 80\% of all superconducting elements have positive Hall coefficients
(meaning hole charge carriers); and d) over 90\% of non-superconducting
metallic, non-magnetic elements have electron charge carriers. This greater
``efficiency'' of holes in producing higher $T_{c}$'s is clearly predicted
in Fig. 4 at least for 2D superconductors, at least insofar as 2h-CP BE
condensates exhibit higher $T_{c}$'s than 2e-CP ones.

\section{Conclusions}

This brief review has sketched how five statistical continuum theories of
superconductivity---including both the BCS and BEC theories---are contained
as special cases in a single theory, the ``complete boson-fermion model''
(CBFM). This model includes, for the first time, both two-electron and
two-hole pairs in freely variable proportions, along with unpaired
electrons, all in chemical/thermal equilibrium. The BCS condensate (characterized by a single equation, namely
the $T$-dependent gap equation) follows directly \textit{as a BE condensate}
through the condition for phase equilibria when both 2e and 2h pair numbers
are equal at a given temperature and coupling---provided the coupling is weak
enough such that the electron chemical potential is roughly the Fermi
energy. Ordinary BEC theory, on the other hand, is recovered from the CBFM
when hole pairs are neglected, fermion-pair coupling is made to vanish, and
the limit of \textit{all} electrons being paired into bosons is taken.

The practical outcome of this BCS-BEC unification via the CBFM is then
threefold: a) \textit{enhancements} in $T_{c}$, by more than an
order-of-magnitude in 2D, and more than two orders-of-magnitude in 3D, are
obtained for the same electron-phonon dynamics mimicked by the BCS model
interaction---provided only that one departs moderately from the perfect
2e/2h-pair symmetry to which BCS theory is intrinsically restricted; b)
these enhancements in $T_{c}$ fall within empirical ranges for 2D and 3D
``exotic'' superconductors, whereas BCS $T_{c}$ values continue to lie low
and within the empirical ranges for conventional, elemental superconductors;
and c) hole-doped superconductors are predicted to have higher $T_{c}$'s
than electron-doped ones, in agreement with observation.\bigskip

\textbf{Acknowledgments }JB and MC are grateful for partial support from
grant PB98-0124 by DGI (Spain). MF, MdeLl\ and VVT thank A.A. Valladares and
H. Vucetich for discussions, and acknowledge UNAM-DGAPA-PAPIIT (Mexico),
grant \# IN106401, and CONACyT (Mexico), grant \# 27828 E, for partial
support. VVT also thanks CONACyT for a chair fellowship at UNAM. Finally,
MdeLl is grateful for travel support through a grant to Southern Illinois
University at Carbondale from the U.S. Army Research Office.

\end{document}